# Phase Transitions of Random Binary Magnetic Square Lattice Ising Systems

**I. Q. Sikakana**
Department of Physics and Non-Destructive Testing,
Vaal University of Technology,
Vanderbijlpark, 1900, South Africa
e-mail: ike@vut.ac.za

**Abstract**

Binary magnetic square lattice Ising system with nearest neighbour interactions were simulated using the Monte Carlo technique. Two types of ions were randomly distributed on the lattice sites, one type interacting ferromagnetic and the other antiferromagnetic. A phase diagram of the ion concentration – dependent critical temperature, $T_c$ was deduced. Combined Bethe Peierls approximation and Mean Field theory phase transition results were compared to the results of the present method. An improved accuracy of the approximations of the critical temperatures was observed. The Monte Carlo simulation is thus shown to be a more reliable method for obtaining the physical properties of the random binary two-dimensional Ising system.

**Keywords:** Binary magnetic two-dimensional Ising system, phase transition, Monte Carlo simulations

## 1. Introduction

The understanding of phase transition properties of magnetic systems is crucial for many applications in engineering and technology [1,2]. The application of the Monte Carlo (MC) simulation method to Ising systems allows one to obtain very accurate physical properties. In previous studies, the XY and Heisenberg models [3]; which consider the potential energy between the spins of unpaired electrons on neighboring atoms were used. Although the individual magnetic moments are small, an extended array of completely ordered moments can generate a large spontaneous magnetization. An important factor in these models is the strength of coupling between the spins, which is sensitive to the interatomic spacings. This then places limitations on the range of interatomic distances suitable for spin coupling. Another important consideration is the magnetic system temperature; however, this is a quantity that is often difficult to define for an isolated magnetic cluster [4], in which the conserved quantity is total energy and not temperature.

Phase transition critical temperature $T_c$, as a function of relative antiferromagnetic coupling ion concentration has been widely studied using the mean field theory. The mean field solutions overestimate the physical properties [5].

A two-dimensional binary Ising square lattices possessing competing nearest neighbor couplings is simulated. The periodic boundary condition (PBC) was adopted for the square lattice, with 60 ions on a side. The system consisted of two types of magnetic ions; one for ferromagnetic coupling, denoted A, and the other for antiferromagnetic coupling denoted B; both randomly distributed on the lattice sites. Magnetization, susceptibility and heat capacity were the properties obtained for different B-ion concentrations. The phase transition critical temperature $T_c$, as a function of relative B-ion concentration was determined.

## 2. Method of Simulation

The Ising Hamiltonian of the system with no external field was defined as:

$$H = -\sum \left[ J_{AA} \xi_{iA} \xi_{jA} + J_{BB} \xi_{iB} \xi_{jB} + J_{AB} (\xi_{iA} \xi_{jB}) \right] S_i S_j, \tag{1}$$

where **J** is the nearest neighbor exchange coupling constant. We define **J** > 0 as ferromagnetic and **J** < 0 as antiferromagnetic. The values used in this work were, respectively, $\{J_{AA}, J_{AB}, J_{BB}\} = \{4, 6, -9\}$ [6]; chosen so that the phase diagram would not be symmetric. Further, for the $i^{th}$ spin we have $S_i = {}^+1$ for spin up↑ and $S_i = {}^-1$ for spin down↓. The type of ions interacting defined as $\xi_{jA} = 1$ for j = A and $\xi_{jA} = 0$ for j = B, $\xi_{jB} = 1$ for j = B and $\xi_{jB} = 0$ for j = A.

To simulate an antiferromagnetic state, the lattices were divided into two interspersing sublattices (checkered), depicted in figure 1 for a 6 by 6 lattice.

```
x o x o x o
o x o x o x
x o x o x o
o x o x o x
x o x o x o
o x o x o x
```

Figure 1. Square lattice decomposed into two B-ion sublattices, with an o-sublattice and x-sublattice

The standard Monte Carlo algorithm, developed by Metropolis et al [7] was used to generate successive magnetic states for a fixed value of $kT/J_{AA}$.

The MC data was sampled as follows; after a start of initial ferromagnetic (FM) or antiferromagnetic (AFM) arrangements, 500 Monte Carlo Steps (MCS) per spin site were discarded, and a further 5000 MCS were simulated at each temperature step. After every 250 MCS, subaverages were determined for that group of states – *termed coarse graining* – to keep track of any major changes in the system. Both heating and cooling sweeps for a particular B-ion concentration were simulated. The physical properties, magnetization $M_\sigma$, susceptibility $X_\sigma$, and heat capacity C, were calculated as follows:

$$|M_\sigma| = \left| \sum S_i \right| / N_\sigma, \tag{2}$$

$$X_\sigma / N = \left[ \langle M_\sigma^2 \rangle - \langle M_\sigma \rangle^2 \right] / kT, \tag{3}$$

$$C/N = \left[ \langle H^2 \rangle - \langle H \rangle^2 \right] / k^2 T^2, \tag{4}$$

With **σ** representing:
  i) the whole system lattice sites, or
  ii) A-ion lattice sites only, or
  iii) B-ion x-sublattice sites only or
  iv) B-ion o-sublattice only.

The absolute values being used for magnetization, since the external field is zero hence no preferred spin direction. The symbol $\langle ... \rangle$ defines an average over N's. The following two conditions are also satisfied:

$$N_A = (1 - P_B) N, \quad (5a)$$

$$P_B = N_{Bx} + N_{Bo}, \quad (5b)$$

where N denotes total number of ions in the lattice, $N_A$ the number of A-ions, $N_{Bx}$ the number of B-ions in x-sublattice and $N_{Bo}$ the number of B-ions in o-sublattice. $P_B$ is the B-ion concentration.

To determine the critical temperature ($T_c$) as a function of relative concentration, at least one of the following methods whose relations hold only near $T_c$, was utilized:

i) sharp onset of magnetization

$$|M| \propto L^{-\frac{\beta}{\nu}} \left[ \varepsilon L^{\frac{1}{\nu}} \right] \quad (6)$$

ii) susceptibility cusp

$$X \propto L^{\frac{\gamma}{\nu}} \left[ \varepsilon L^{\frac{1}{\nu}} \right] \quad (7)$$

iii) heat capacity peak

$$C \propto L^{\frac{\alpha}{\nu}} \left[ \varepsilon L^{\frac{1}{\nu}} \right] \quad (8)$$

iv) finite size scaling analysis

With $\varepsilon = (T - T_c)/T_c$, and the critical parameters for the two dimensional Ising square lattice being; $\nu = 1$, $\gamma = \frac{7}{4}$, $\beta = \frac{1}{8}$ and $\alpha \approx (\log \varepsilon)$ [8].

Initially, at least ten different ion distributions were produced for a given relative B-ion concentration. For each ion distribution, the cluster sizes of thirty (30) ions or more were noted. The most representative distribution was identified and used in the calculations at that concentration. A sample distribution for the B-ion concentration, $P_B = 0.55$ is shown in table 1. The importance of this exercise was to build confidence in the properties deduced from the system, that is, they should be comparable with those in a very large system.

Table 1.     B-ion concentration, $P_B = 0.55$ with ten different ion distributions; for clusters of at least thirty (30) ions or more.

| Distribution Number | A – ion Cluster sizes | B – ion Cluster sizes |
|---|---|---|
| 1 | 75,72,59,40,38,36,34,33,32,32 | 534,288,183,171,80,46,44,41,34,34 |
| 2 | 65,42,40,39,35,32,32,30 | 841,346,67,48,36,31 |
| 3 | 135,85,50,46,31,30,30 | 411,409,205,96,77,64,58,48,43,35 |
| 4 | 75,61,57,54,50,47,38,36,33,33 | 412,280,248,94,86,62,56,50,46,34 |
| 5 | 99,53,49,48,44,41,41,39,36 | 328,218,149,143,103,92,81,67,60,39 |
| 6 | 78,71,65,49,44,39,37,34 | 198,189,184,143,135,130,102,95,76,71 |
| 7 | 70,65,60,50,42,39,38,36,34,31 | 782,189,122,92,76,71,68,59,52,42 |
| 8 | 116,61,60,47,46,45,34,34,33,32 | 454,394,359,188,50,33 |
| 9 | 83,69,60,48,48,46,38,36,36,32 | 833,266,107,92,77,71,57,56,35 |
| 10 | 79,50,46,44,42,42,40,33,31,31 | 742,191,191,145,103,70,38 |

## 3. Results and Discussions

### 3.1 High and Low B-ion concentration

The ranges $0.70 < P_B < 1.00$ for high B-ion concentration and $0.00 \leq P_B \leq 0.20$ for low B-ion concentration were simulated. In obtaining the critical temperature $T_c$, finite size scaling analysis for a system with PBC was utilized. Figures 2 and 3 shows the log-log plots of $M L^{\frac{1}{8}}$ vs. $\varepsilon L$ for the low (figure 2) and the high (figure 3) B-ion concentrations. For each B-ion concentration, a straight line through the points was fitted using the linear least-squares fit method. The best estimate for $T_c$ was taken to be that value for which the slope of the fitted line was closest to $\frac{1}{8}$.

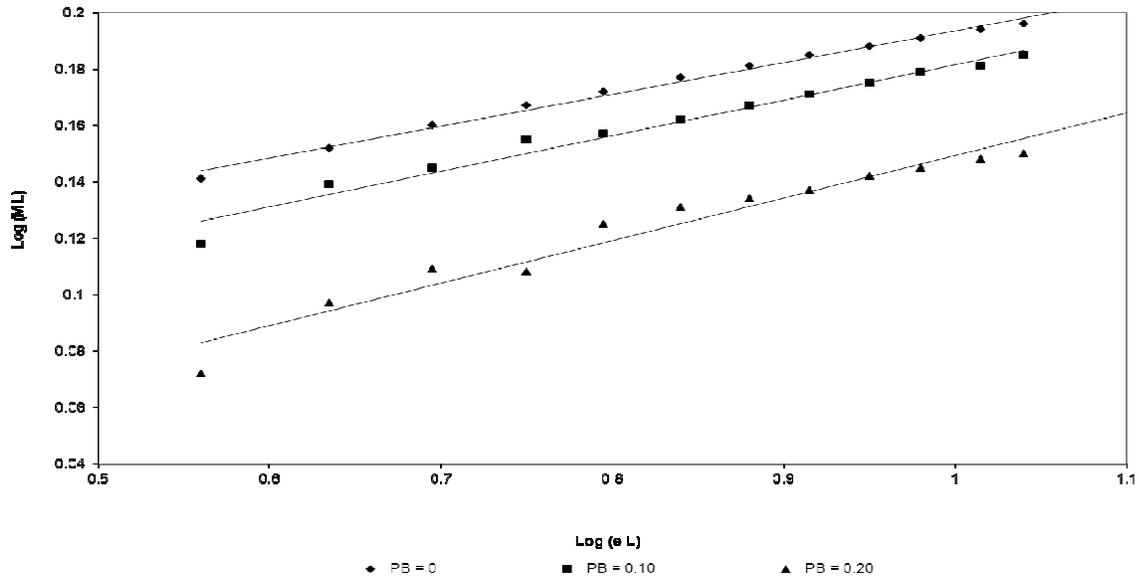

Figure 2.     Plots of Log $(ML^{\frac{1}{8}})$ versus Log $(\varepsilon L)$ for $P_B = 0.00$, 0.10 and 0.20. Finite size scaling analysis used with the slopes of the lines being 0.125±.

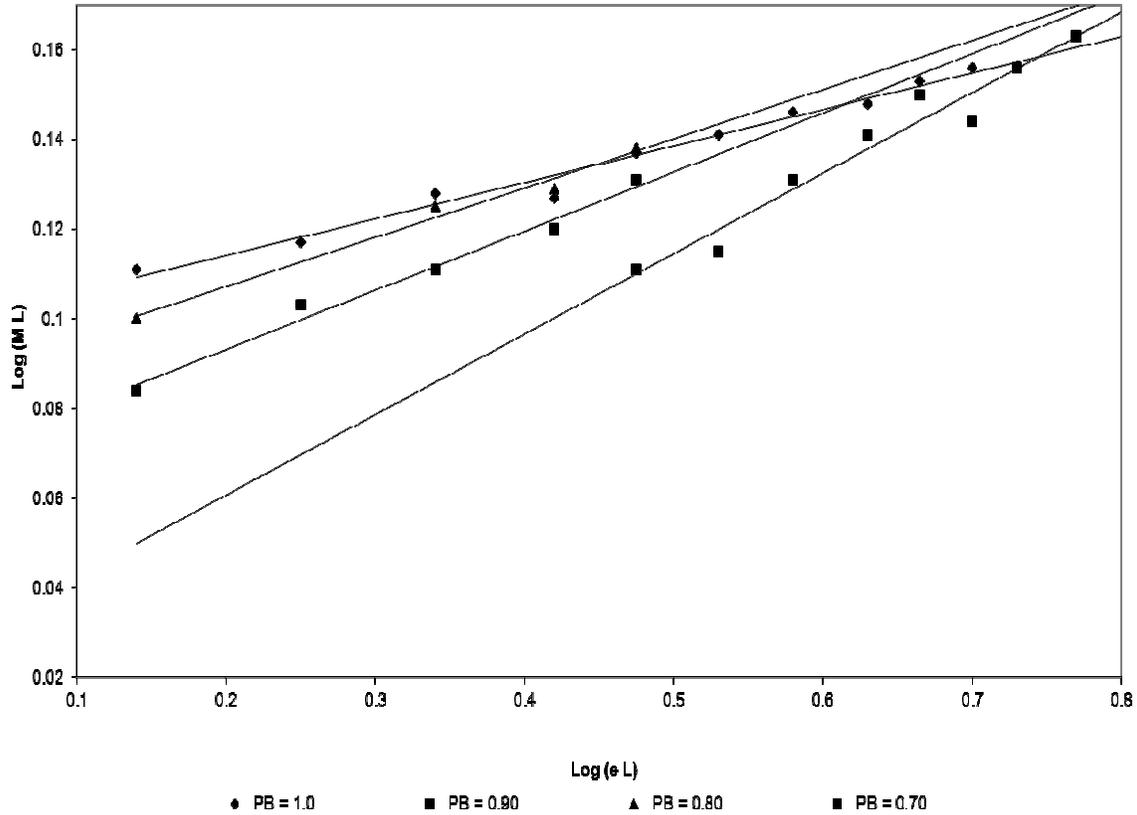

Figure 3.   Plots of Log (ML$^{1/8}$) versus Log ($\varepsilon$L) for $P_B$ = 0.70, 0.80, 0.90 and 1,0, Finite size scaling analysis used the slopes of the lines ranging from 0.2 - 0.125.

## 3.2 Mid B-ion concentration

At these concentrations 0.20 < $P_B$ < 0.70 sweeps of 5000 MCS were found to be inadequate for the following two reasons:
i)   heating and cooling sweeps for the same ion concentration did not coincide with each other.
ii)  the initial state (FM or AFM) of the system was influencing the results.
To improve the results, 50000 MCS were done at each temperature step.  In figure 4 and 5, data for $P_B$ = 0.35 and 0.60 are respectively plotted for both the system and Bx – ions magnetization and susceptibility.  For $P_B$ = 0.35, the magnetization was less than half for both ion distributions shown, which meant that the system did not become ferromagnetic, whilst the susceptibilities showed little agreement.  For $P_B$ = 0.60, both ion distributions produced some antiferromagnetic ordering, with the susceptibilities showing good agreement.

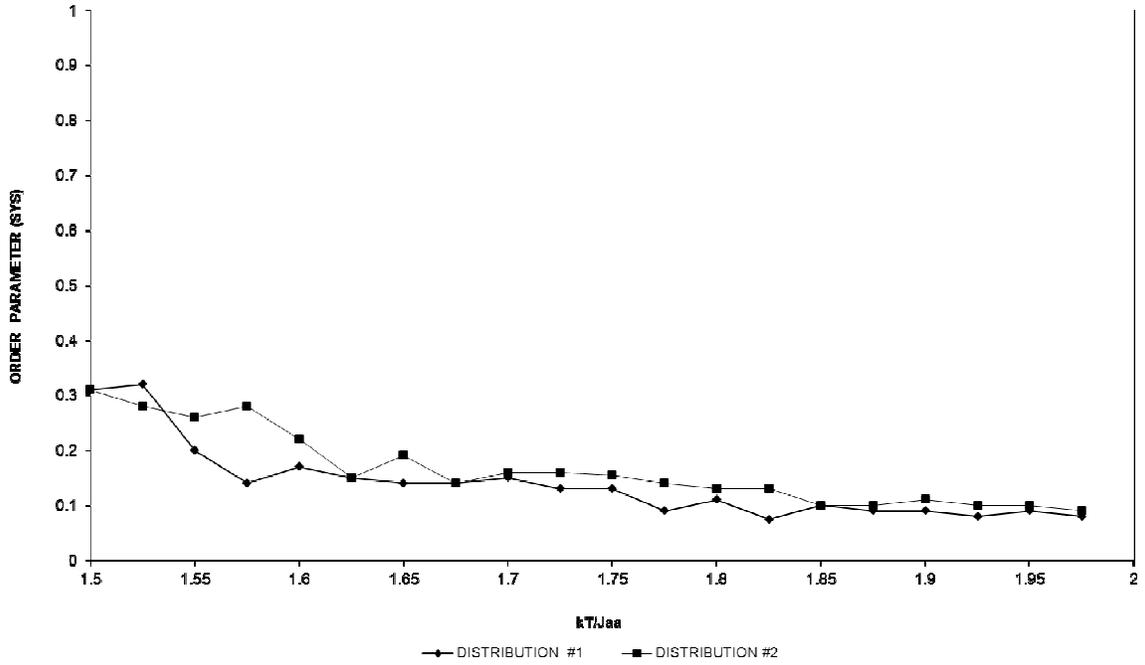

Figure 4a. System order parameter for $P_B = 0.35$ with two ion distributions.

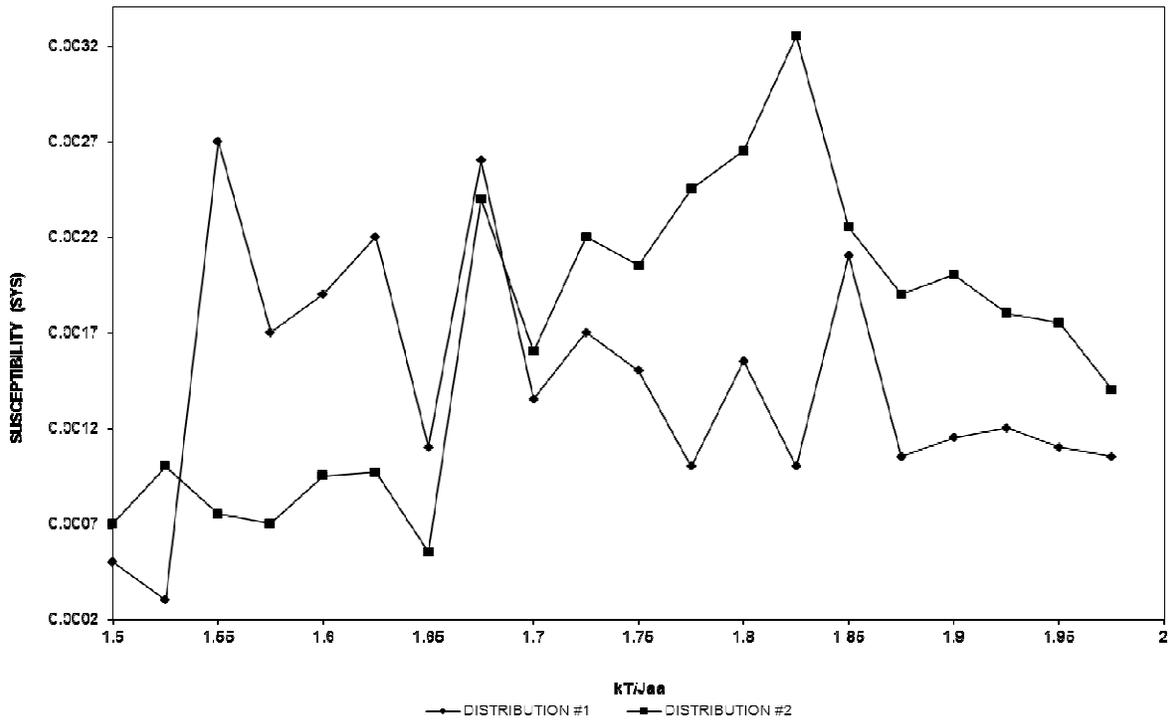

Figure 4b. System susceptibility for $P_B = 0.35$ with two ion distributions.

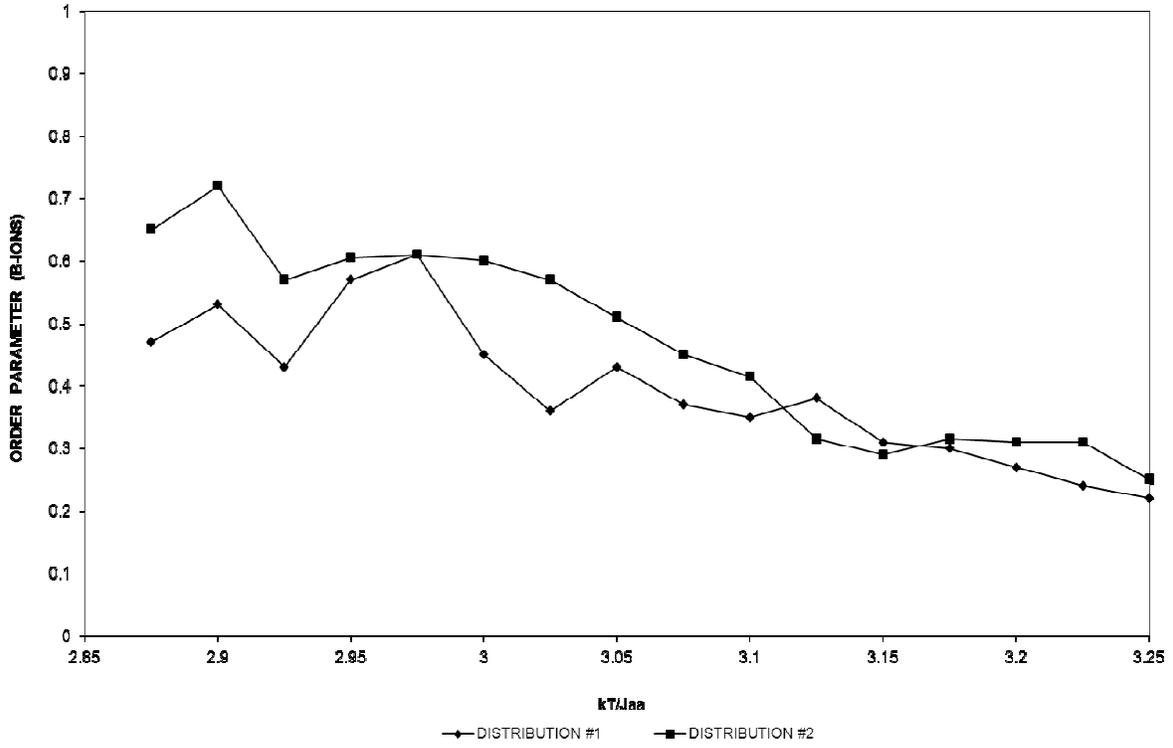

Figure 5a. $B_x$–ions order parameter for $P_B = 0.60$ with two ion distributions.

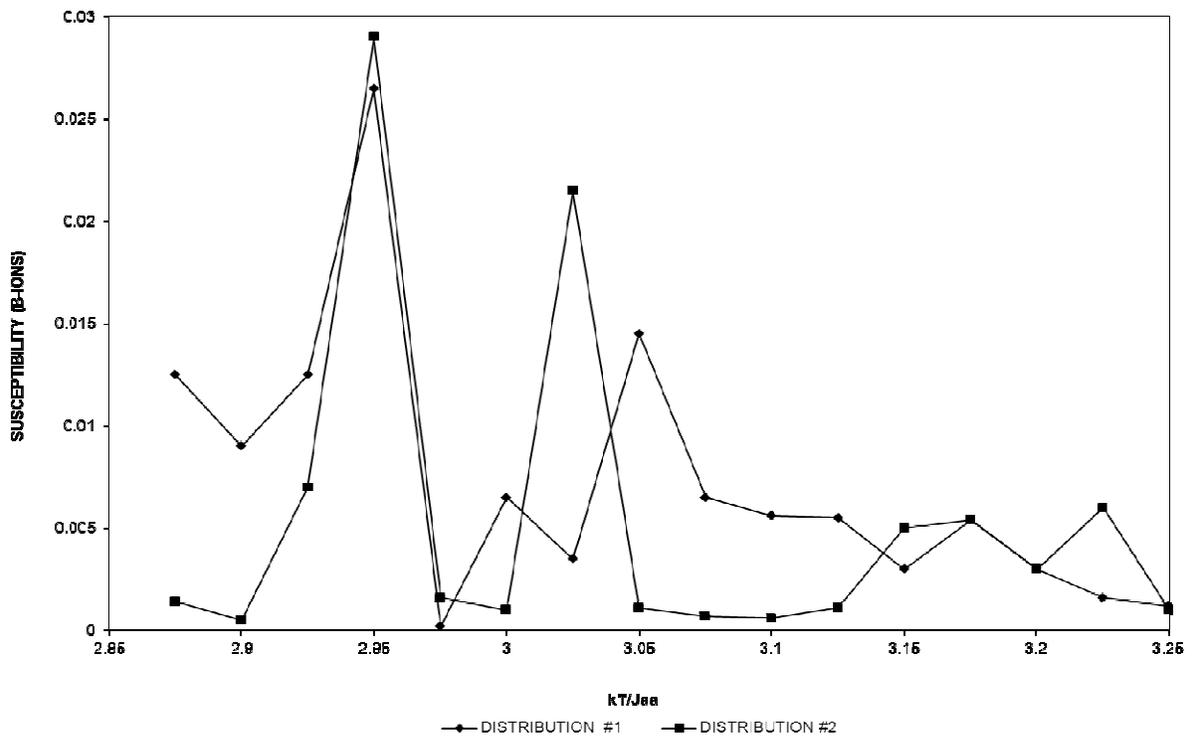

Figure 5b. $B_x$–ions susceptibility for $P_B = 0.60$ with two ion distributions.

Systems where the ion concentration was such that neither of the two types of ions had ordered clusters spanning the system were simulated. According to percolation theory [9], no long-range phase transitions were expected, rather short-range ordering should be evident. Figure 6, shows the plot for $P_B = 0.45$. Very little variation in the magnetization was observed, even at low temperatures, whilst susceptibility showed restricted variations with no real cusp.

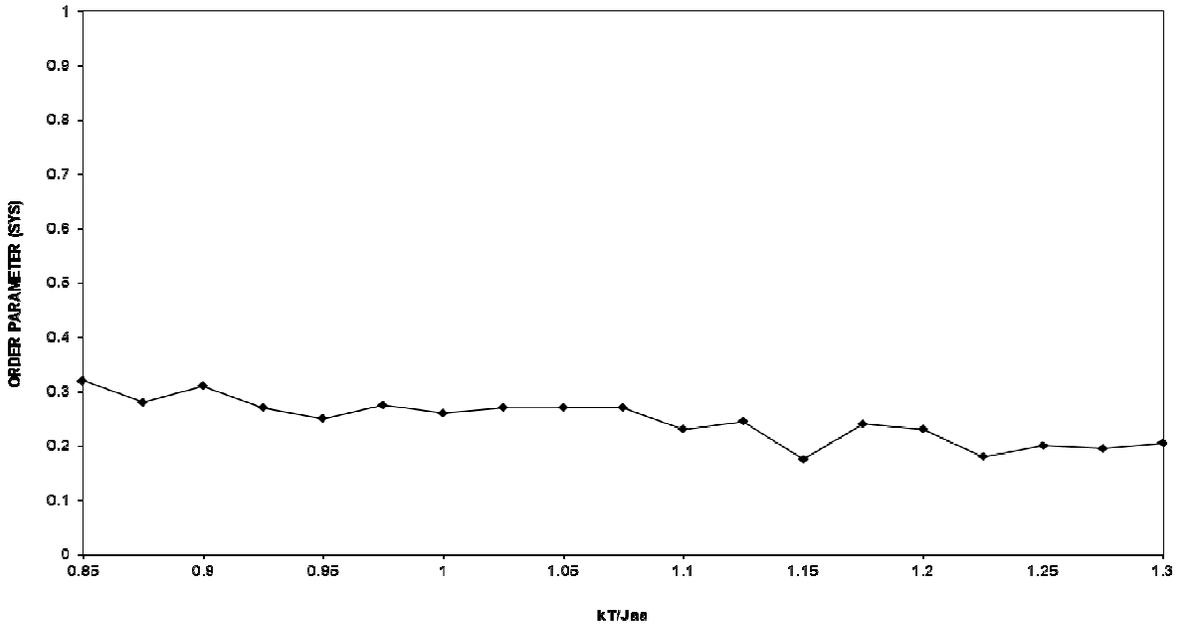

Figure 6a. System order parameter for $P_B = 0.45$ with one ion distribution.

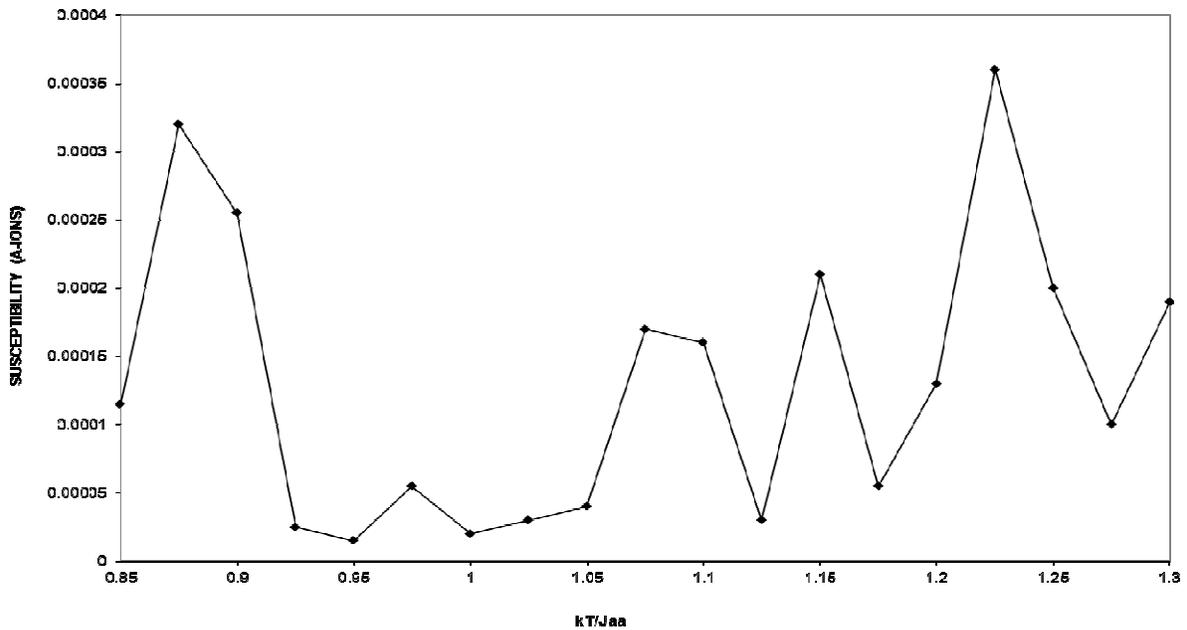

Figure 6b. A – ions susceptibility for $P_B = 0.45$ with one ion distribution.

A *quench test* was then done, since there was evidence that as the system was cooled in small temperature steps, the B-ions froze in phase and prevented the A-ions from ordering. The quench test involved starting the system at high enough temperature so that after a number of MCS, a paramagnetic state was achieved, immediately thereafter one large temperature decrease was taken followed by small temperature steps. Suffice to report that barely any change in the results was observed.

### 3.2.1 Frustration and A-B couplings

As the estimation of $T_c$ at these B-ion concentrations has large errors, another investigation was done to find the influence of the A-B coupling, by simulating similar systems with the $J_{AB}$ interaction switched off (set to zero). From the results found, it was concluded that at these concentration ranges the A-B coupling was partly responsible for the frustration in the system.

### 4. Phase Diagram

The phase transition temperatures of all the simulated ion concentrations were compiled and a phase diagram was plotted as shown in figure 7. The uncertainties in $T_c$'s were found using the following two methods:
i)   Standard error of the slope recorded by linear regression for finite size scaling analysis.
ii)  Half-width at half maximum of the peak from susceptibility curves.

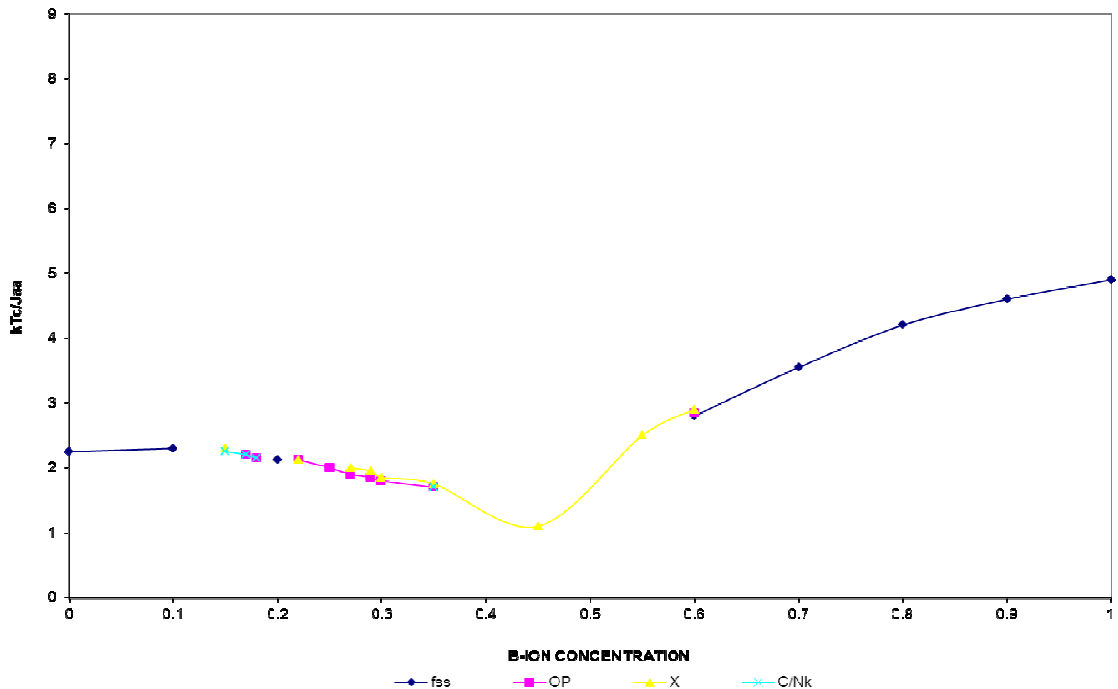

Figure 7. Phase diagram for ferromagnetic – antiferromagnetic random site model with nearest neighbour interaction. The blue solid line is the second-order phase transition, and the other coloured lines are the short-range order transitions. Different marked points are for finite size scaling analysis (fss), order parameter (OP), susceptibility (X) and heat capacity (C/Nk) results.

The uncertainties for the critical temperatures determined from magnetization and heat capacity, were found to be of similar magnitude as evident from the phase diagram. Further comparison with $T_c$'s found using susceptibility shows an increase in the uncertainties.

A paper on random site binary magnetic Ising systems using a Bethe-Peierls approximation and mean field theory [10], concluded that depending on the strength of the coupling between the two types of ions, two kinds of phase diagrams are possible:

i) Strong coupling between the ions $-J_{AA} \rangle \sqrt{J_{AA}/J_{BB}}$; the ferromagnetic and antiferromagnetic regions were separated by a line of first-order transition.

ii) Weak coupling between the ions $-J_{AB} \rangle \sqrt{J_{AA}/J_{BB}}$; the ferromagnetic and antiferromagnetic regions were separated by a mixed phase in which both types of order coexisted.

The mean field equations were applied to the present work and the deduced phase diagram is shown in figure 8.

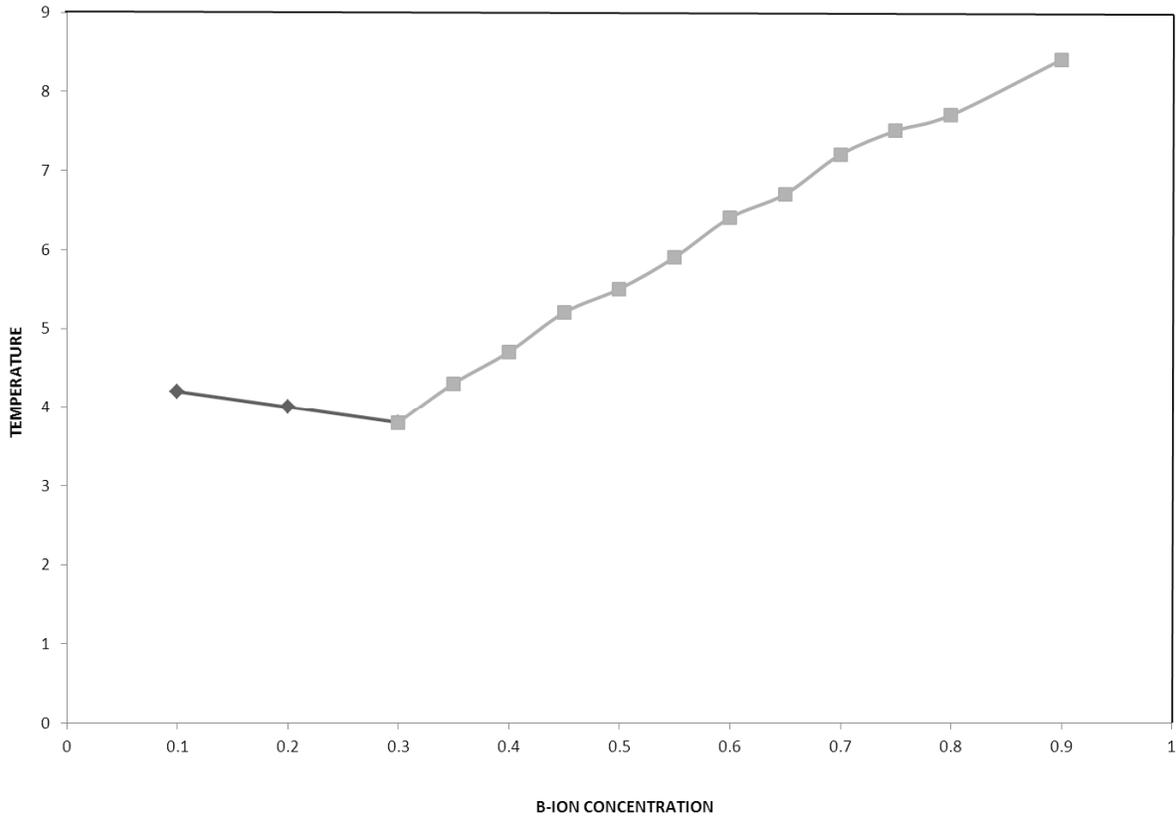

Figure 8. Phase diagram calculated using the mean field equations of Eggarter & Eggarter [10], with the nearest neighbour interactions; $\mathbf{J}_{AA} = 1$, $\mathbf{J}_{AB} = 1.5$ and $\mathbf{J}_{BB} = -2.25$.

In comparison to the Monte Carlo results (figure 7), there is a significant difference in the concentration dependent transition temperatures. This can be ascribed to the fact that the mean field results over estimates the critical temperature.

## 5. Conclusion

The present Monte Carlo simulations of the two-dimensional binary Ising system determined the critical temperatures $T_c$, accurately for high and low B-ion concentrations.

The determination of critical temperatures in the mid B-ion concentration range proved to be difficult, even after the number of MCS was increased tenfold. After analysis of the results, it can be deduced that the B-B coupling had a lot to do with this frustration. This could have been because the B-B coupling was preventing A-ions from ordering or making it possible only for short-range ordering of the B-ions. The influence of the A-B coupling was investigated and the findings were inconclusive. In the percolation region ($0.41 \leq P_B \leq 0.59$), no true phase transitions were identified, but only short-range order was observed.

The overall character of the phase diagram shows improvement in terms of system transition temperature accuracy. Comparison to the mean field theory phase diagram revealed the superiority of the present method over analytical theory calculations.

**Acknowledgement**

I would like to thank Professor P D Scholten (Department of Physics, Miami University, Oxford, Ohio, USA) without whom this work would not have been possible.